\documentclass[a4paper,11pt]{article}
\usepackage{mathrsfs}
\usepackage{amssymb}
\usepackage{stmaryrd}
\usepackage{txfonts}
\usepackage[a4paper,left=3cm,right=3cm]{geometry}
\usepackage{graphicx}

\newcommand{\be}{\begin{equation}}
\newcommand{\ee}{\end{equation}}
\newcommand{\bea}{\begin{eqnarray}}
\newcommand{\eea}{\end{eqnarray}}

\usepackage{authblk}

\begin{document}
\setlength{\unitlength}{1.0mm}
\title{The dynamical transition in proteins and non-Gaussian behavior of low frequency modes in Self Consistent Normal Mode Analysis}

\author[1]{Jianguang Guo}
\author[2]{Timo Budarz}
\author[3]{Joshua M. Ward}
\author[1]{Earl W. Prohofsky\thanks{ewp@purdue.edu}}
\affil[1]{Department of Physics, Purdue University, West Lafayette,
IN 47907}

\affil[2]{Department of Physical Science, Santa Ana College, Santa
Ana, CA 92706}

\affil[3]{Department of Medicinal Chemistry and Molecular
Pharmacology, West Lafayette, IN 47907}
\date{}
\maketitle \normalsize
\begin{abstract}
Self Consistent Normal Mode Analysis (SCNMA) is applied to heme c
type cytochrome f to study temperature dependent protein motion.
Classical Normal Mode Analysis (NMA) assumes harmonic behavior and
the protein Mean Square Displacement (MSD) has a linear dependence
on temperature. This is only consistent with low temperature
experimental results. To connect the protein vibrational motions
between low temperature and physiological temperature, we have
incorporated a fitted set of anharmonic potentials into SCNMA. In
addition, Quantum Harmonic Oscillator (QHO) theory has been used to
calculate the displacement distribution for individual vibrational
modes. We find that the modes involving soft bonds exhibit
significant non-Gaussian dynamics at physiological temperature,
which suggests it may be the cause of the non-Gaussian behavior of
the protein motions probed by Elastic Incoherent Neutron Scattering
(EINS). The combined theory displays a dynamical transition caused
by the softening  of few "torsional" modes in the low frequency
regime ($<50cm^{-1} or <6meV or
>0.6ps$). These modes change from Gaussian to a classical distribution upon heating. Our theory provides an
alternative way to understand the microscopic origin of the protein
dynamical transition.
\end{abstract}

\section{Introduction}
Protein function is determined by both structural stability and
flexibility. The stability is needed to ensure appropriate geometry
of the protein, while the flexibility allows function to proceed at
an appropriate rate. Quantitative measurements of the temperature
dependent atomic mean square displacements (MSD) are possible by
neutron scattering \cite{Doster_exp1} \cite{Doster_exp2}\cite
{Zaccai1}and M\"ossbauer
absorption\cite{Parak_exp1,Parak_exp2,Parak_exp3}. All of these
experiments show a "dynamical transition" in hydrated proteins,
which is marked by an abrupt MSD increase in the temperature range
160--240K. It is believed that this dynamical transition is
correlated with protein function. Three prominent examples are the
myoglobin-CO binding kinetics\cite{myoglobin-co}, electrostatic
relaxation in green fluorescent protein\cite{Green}, and the
Arrhenius behavior of the electron transfer rate above the dynamical
transition temperature. However, the time scale and the forms of the
functionally important atomic modes remain a subject of active
discussion\cite{Doster_discussion}\cite{Ringe}\cite{Sokolov}.

Numerous theoretical studies of protein dynamics have been carried
out by molecular dynamics (MD) simulations
\cite{Smith_MD1,Smith_MD2,Smith_MD3,Smith_MD4} and normal mode
analysis (NMA)
\cite{Smith_MD3,Go_1983,Brooks_1983,Parak_NMA,Kneller}. NMA requires
the use of Maxwell-Boltzmann or Gaussian distributions to describe
the probability distributions of individual atoms or chemical bonds.
Recently, several authors focused on the study of the non-Gaussian
behavior of the total elastic incoherent neutron scattering (EINS)
profile from a protein above dynamical transition
temperature\cite{Doster_gaussian1,Kataoka1,Kataoka4}. It should be
noted that the distribution of all-atom MSDs from an EINS profile
can still be non-Gaussian even if all atoms individually exhibit
Gaussian dynamics. The Gaussian distribution, which is the ground
state probability distribution for the quantum harmonic oscillator,
is an appropriate approximation when $\hbar\omega>kT
(\omega>~200cm^{-1}$). In the Gaussian distribution, the atom has
maximum probability in the equilibrium position. We find that in all
self consistent theories, the use of a Gaussian distribution results
in a molecular structure that will tend to be more rigid than what
would be found by a more exact quantum approach.  From Newton's
second law, the classical harmonic oscillator (low frequency) has
highest probability at the edges of the well because the atom moves
most slowly near the classical turning points,  which is contrary to
the Gaussian or ground state probability distribution. The exact
quantum behavior of low frequency modes would approach the classical
displacement. In this paper we explore the role of incorporating the
higher quantum vibrational states. This shows a softening of the
structure in the correct temperature range.

The material studied by SCNMA is six-coordinate heme c type
cytochrome f\cite{cytf}. The iron normal modes are compared with the
Nuclear Vibrational Resonance Spectroscopy (NRVS) spectrum
\cite{NRVS}. NRVS is uniquely capable of displaying the low
frequency vibrational displacement spectrum of the Fe atom at the
center of the heme as it sees all modes and can give quantitative
values for displacements. It is then possible to define low
frequency heme modes that are in agreement with observation with
greater accuracy. This is a much more stringent test than most Raman
comparisons as Raman displacements cannot be calculated with any
accuracy.

SCNMA incorporates non-linearity into harmonic calculation by
thermal-statistically averaging the curvature of the bond potential
energies. Because vibrational modes that are not over-damped are
detected by Raman and IR, one expects the effective Hamitonian to be
approximately harmonic. SCNMA should therefore be a valid approach.
The SCNMA formulation arises from a variational procedure that finds
the best effective harmonic Hamiltonian by minimizing the free
energy. This method is described in detail
elsewhere\cite{Prohofsky}. It has been successfully employed on
models with multiple hydrogen stretching bonds such as the helix
melting, conformational change in DNA and drug-helix stability, etc
\cite{Prohofsky_SCNMA1}-\cite{Prohofsky_SCNMA4}. In those papers a
Gaussian distribution was used to describe the displacement
distribution. In this paper, we will further develop this method to
incorporate non-Gaussian distributions into our calculation.

\section{Quantum Harmonic Oscillator (QHO) theory applied to internal atomic bonds}

\subsection{The displacement distribution of the internal atomic
bonds}
For a bio-molecule with N atoms and M internal atomic bonds M
is much larger than N. Standard NMA will give us 3N-6 non-zero
normal modes. Their frequencies can be written as
$\omega=[\omega_1,\omega_2, ...,\omega_{3N-6}]$. The total MSD for
frequency $\omega$ can be written as
\be <\sum_{i=1}^nm_ir_i^2>=\frac{\hbar}{2\omega}coth(\frac{\hbar
\omega}{2k_BT})  \ee

Subsequently, the temperature dependent total mean square amplitude
for the one single internal bond is the sum of all normal mode
amplitudes, which can be written as \be D^2=\sum_\omega
D_\omega^2=\sum_\omega d_\omega^2
coth(\frac{\hbar\omega}{2k_BT})=\sum_\omega
coth(\frac{\hbar\omega}{2k_BT})|{s_\omega}|^2 \ee

 where $D^2$ is the total mean square amplitude over all frequency modes,
$D_\omega^2$ is the mean square amplitude contribution and
$d_\omega^2$ is the zero point mean square amplitude for frequency
$\omega$, and $|s_\omega|^2$ is the projection of the normalized
eigenvectors at eigenvalues (frequency) $\omega$ onto the
mass-weighted internal coordinates. These amplitudes can represent a
linear distance (for stretching bond) or an angular twisting (for
angle bend and dihedral bond).

From QHO theory, the harmonic displacement distribution for one
particular internal bond at frequency $\omega$ can be written as \be
<u_\omega|H|\Psi_n>=\sqrt{\frac{1}{2^n n!}}(\frac{1}{2\pi
d_\omega^2})^{1/4} e^{-\frac{u_\omega^2}{4d_\omega^2}}H_n
(\sqrt{\frac{1}{2 d_\omega^2}}u_\omega) \ee where $u_\omega$ is the
displacement variable for mode $\omega$, $H_n$ is the Hermite
polynomial, and $\Psi_n$ is the probability distribution for the
$n^{th}$ excitation state. Here we note that the ground state $
<u_\omega|H|\Psi_0>$ is in fact a Gaussian. The corresponding
quantized energy levels are \be E_n=\hbar\omega(n+\frac{1}{2}) \ee
From the Boltzmann distribution, the displacement distribution of
this internal coordinate for mode $\omega$ can be written as \be
f_\omega(u_\omega)=\frac{\sum_{n=0}^\infty
e^{\frac{-\hbar\omega(n+\frac{1}{2})}{kT}<u_\omega|H|\Psi_n>^2}}{\sum_{n=0}^\infty
e^{\frac{-\hbar\omega(n+\frac{1}{2})}{kT}}} \ee The joint
probability density function for $\omega=[\omega_1,\omega_2,
...,\omega_{3N-6}]$ can be subsequently written as \be
g(u_{\omega_1},u_{\omega_2},...)=\prod_\omega f_\omega (u_\omega)
\ee The total displacement is $u=\sum_\omega u_\omega$. Using a
transformation of variables \be [u=\sum_\omega u_\omega,
u_{\omega_2}=u_{\omega_2}, u_{\omega_3}=u_{\omega_3},...]    \ee the
total displacement distribution can be obtained as \be
f(u)=\int_{-\infty}^\infty\cdot\cdot\cdot\int_{-\infty}^\infty
g(u-\sum_{j=2}^{3N-6}u_{\omega_j},u_{\omega_2},u_{\omega_3},...)du_{\omega_2}du_{\omega_3}\cdot\cdot\cdot
u_{\omega_{3N-6}}   \ee

To reduce unsystematic errors, $u_{\omega_1}$ is chosen to have the
largest amplitude of the $3N-6$ modes. Equation 8 requires $3N-7$
integrals of a $3N-6$ multi-variable function to calculate the
actual displacement distribution of one single internal bond. For
one standard NMA calculation, $(3N-7)\times M$ integrals are solved.
To reduce the required calculation time, approximation methods are
employed, as introduced in the next section.

\subsection{The displacement distribution for single frequency harmonic motion and an approximation method}

To understand the approximate temperature and frequency behavior of
non-Gaussian distributions, that of a single frequency normal mode
displacement is shown in Fig 1.  It shows the temperature dependent
single frequency displacement distribution at 300K for
(a) $\omega<50cm^{-1}$ $(>0.67ps)$,
(b) $50cm^{-1}<\omega<80cm^{-1}$ (0.42ps-0.67ps), and
(c) $\omega>80cm^{-1}$.

\begin{figure}[h]
\begin{center}
\includegraphics[width=14cm]{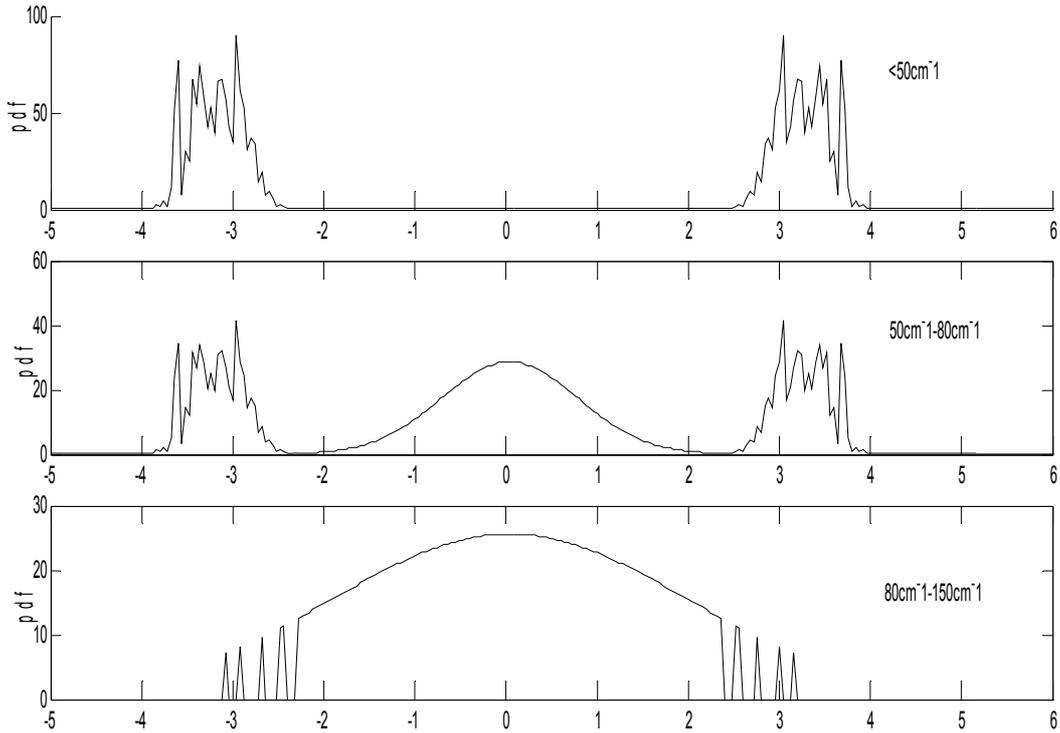}
\caption{ Characterization of the single frequency displacement
distribution at 300K In (a) the displacement is similar to the
classical distribution, (c) the displacement distribution is similar
to a Gaussian, and (b) is a cross between the two}
\end{center}
\end{figure}

Fig 1 shows the displacement probability for a single frequency, but
the displacement for a single bond is a superposition of many such
frequency contributions with different amplitudes.  The spread in
amplitudes comes from the projection factors ($|s_\omega|^2$) from
equation $(5)$ which come from the eigenvectors of the various
modes.  Even for low frequencies, any bond amplitude would be the
sum of many distributions like those in Fig 1, all at different
amplitudes from the origin.  The central Limit Theorem (CLT)
supposes that a large sum of this kind will add up to a Gaussian
distribution. This assumption has been central to all previous
calculations using SCNMA. The situation could be quite different,
however, if only a few low frequency modes dominate in the
displacement of particular bonds. In such a case, for some range of
temperatures, the displacement probability could resemble the plot
in Fig 1-(a).   We emphasize that the hydrogen bond stretching modes
are typically above $100cm^{-1}$ and fall into the Gaussian
distribution regime. The bond modes that are softer than the
hydrogen stretching bonds, i.e. the torsional motions, may exhibit
non-Gaussian behavior at physiological temperature. All proteins
have torsional modes and this effect may be manifested in many
proteins.

From equation (5), the displacement distribution of the single
frequency mode is approximately Gaussian when \be
\omega(T)>0.27Tcm^{-1}   (T in Kelvin)   \ee and more classical when
\be \omega(T)<0.17Tcm^{-1}   (T in Kelvin) \ee From equations (9)
and (10), the single frequency mode in the frequency regime
$<50cm^{-1}$ (or $<6 meV$ or $>0.7 ps$) will transition from a
Gaussian to a more classical distribution upon heating from low
temperature to room temperature. It should be noted that the
prominent "Boson peak" $(1-3.5meV or 10-30cm^{-1})$ from neutron
scattering \cite{Doster_mode1, Zaccai1, Frick} or the "doming mode"
from NRVS \cite{NRVS} and IR \cite{doming_exp2} experiments lie in
this frequency regime.

To simplify the calculation, we use the assumption that the sum of
the independent Gaussian variables is still a Gaussian and we treat
all the normal modes above $80cm^{-1}$ as one Gaussian distribution.
Based on CLT, we can further simply the low frequency displacement
distribution calculation. If the displacement $u$ for one internal
bond is comprised of many low frequency modes, we can treat it as a
Gaussian. To test how many significant low frequency $(<80cm^{-1})$
modes are needed to be able to use the Gaussian approximation
without loss of accuracy,  several NMA and subsequent displacement
distribution calculations were run on the heme core. We found less
than $5\%$ deviation from Gaussian in the distribution of $u$
(equation 8) when $u$ has more than 5 low frequency modes each
accounting for more than $10\%$ of the total potential energy.
Implementing these two approximations reduce our calculation time by
a factor of more than 100.

\section {Method}
An initial classical NMA calculation was performed on the six-coordinate heme c type
cytochrome f using the CHARMM force field \cite{Karplus_CHARMM,MacKerell_1998}.
An all-atom model \cite{MacKerell_1998} was
constructed from the X-ray coordinates (PDB identifier 1EWH
\cite{cytf}). The model was subjected to force field minimization
until the root mean square gradient of the potential energy was less
than 0.0001 prior to performing a standard normal mode calculation with
the VIBRAN facility in CHARMM\cite{Karplus_CHARMM}.

The low temperature CHARMM force field was refined by comparison
with the Nuclear Vibrational Resonance Spectroscopy (NRVS)
spectrum\cite{NRVS}. The method of force field refinement process
was described elsewhere\cite{Brajesh, Brajesh_NRVS, Timo_NRVS}. The
anharmonic functional forms were chosen from reference \cite{Mayo},
in which Morse function, harmonic cos function and dihedral cos
function were used to describe bond stretch interactions, angle bend
interactions and torsional bond interactions. The resulting low
temperature force constants can then be used along with data on atom
distances and bond strength to fit anharmonic potential parameters.
SCNMA was employed to allow exploration of temperature dependent
changes in force constant and thermal expansion
effects\cite{Prohofsky}. This method has been described in detail
elsewhere\cite{Prohofsky, Prohofsky_SCNMA1, Prohofsky_SCNMA2,
Prohofsky_SCNMA3, Prohofsky_SCNMA4}, where the Gaussian
approximation was used for the displacement variable $u$. The only
difference here from the previous SCNMA is the explicit inclusion of
non-Gaussian distributions for low-frequency modes. Here, we give a
brief description of the computation:

\begin{itemize}
\item Input the effective force constants (the $1^{st}$ iteration
uses the force constants refined to experimental data) into the NMA
and find the initial normal mode eigenvalues and eigenvectors.

\item  Calculate each internal coordinate's total mean square
amplitude $D^2$ and each normal mode contribution $D_\omega^2$

\item Calculate each internal coordinate displacement distribution
$f(u)$

\item Calculate a new set of the effective force constants.

\item Iterate to self consistency.
\end{itemize}
The calculation converged within 20 iterations.

\section {Result and discussion}

Fig 2 shows the comparison between the iron Vibrational Density Of
State (VDOS) obtained from classical NMA and the NRVS experimental
results. Good agreement is achieved over a wide range of
frequencies, which indicates a useful choice for the low temperature
limit force field. Here, we give a summary of the general results:
(1) below $80cm^{-1}$ are mostly iron-out-of-plane motions; (2) the
$80-300cm^{-1}$ region has both iron in-plane and out-of-plane
features; and (3) $>300cm^{-1}$ are mainly iron in-plane motions. If
the calculations did not include anharmonic effects, the total
displacements would be linear in temperature. Classical NMA results
show that at low temperature ($<150 K$), the iron out-of-plane MSD
is about three times the iron in-plane MSD despite the fact that the
iron in-plane motion has two degrees of freedom versus the single
degree of out-of-plane motion.
\begin{figure}[h]
\begin{center}
\includegraphics[width=14cm]{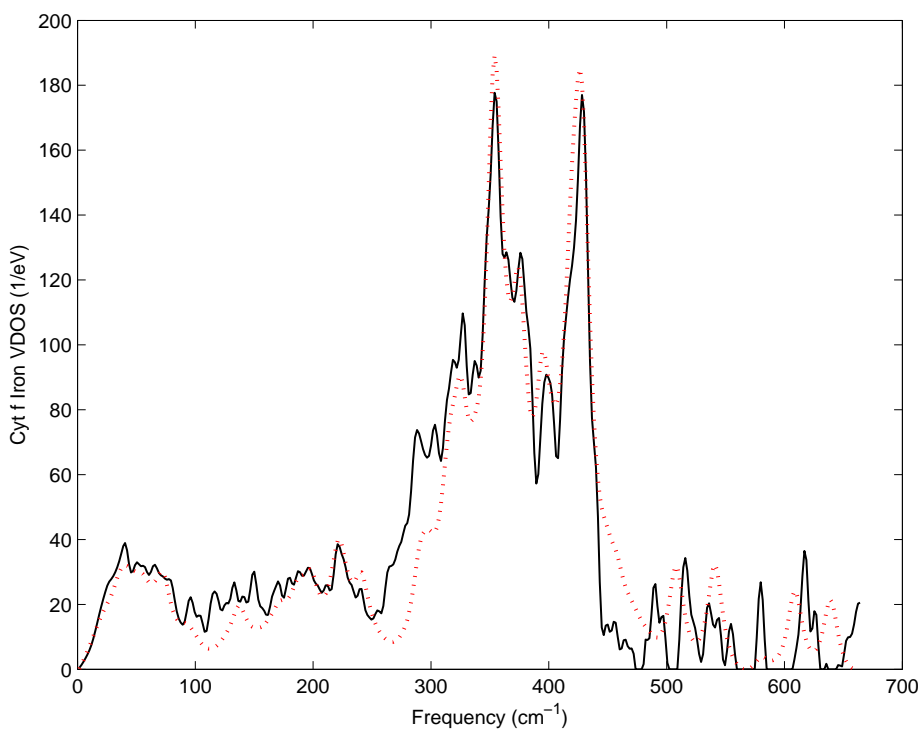}
\caption{ Comparison of the experiment (from reference \cite{Adams})
and theoretical (from classical NMA) cytochrome f iron Vibrational
Density Of State (VDOS) Solid black line: experiment; dotted red
line: theory}
\end{center}
\end{figure}

\begin{figure}[h]
\begin{center}
\includegraphics[width=14cm]{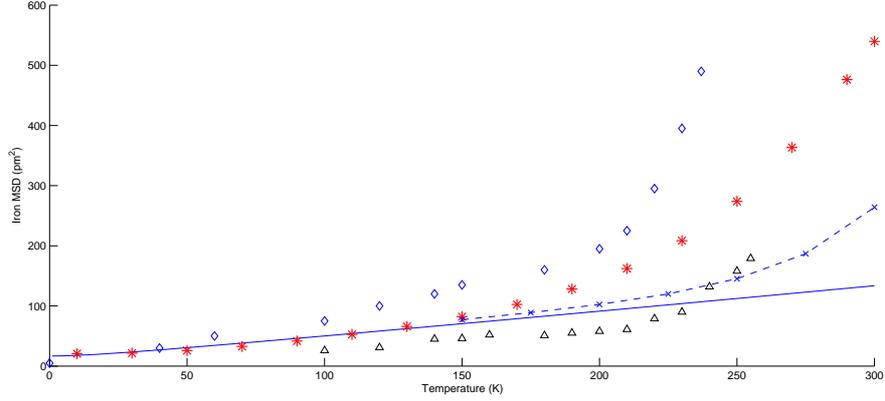}
\caption{ The iron MSD vs. temperature plot for various heme
proteins. Blue line: iron MSD of cyt f by classical NMA; Blue cross
and blue dashed line: iron MSD of cyt f from SCNMA with Gaussian
distribution approximation. Red star: iron msd of cyt f from our
SCNMA by implementing non-Gaussian displacement distribution. Blue
diamond: iron MSD of myoglobin by M\"ossbauer absorption measurement
from ref. \cite{Parak_exp1}. Black upper triangle: iron MSD of cyt c
by M\"ossbauer absorption measurement from ref. \cite{Parak_exp2}}
\end{center}
\end{figure}

\begin{figure}[h]
\begin{center}
\includegraphics[width=14cm]{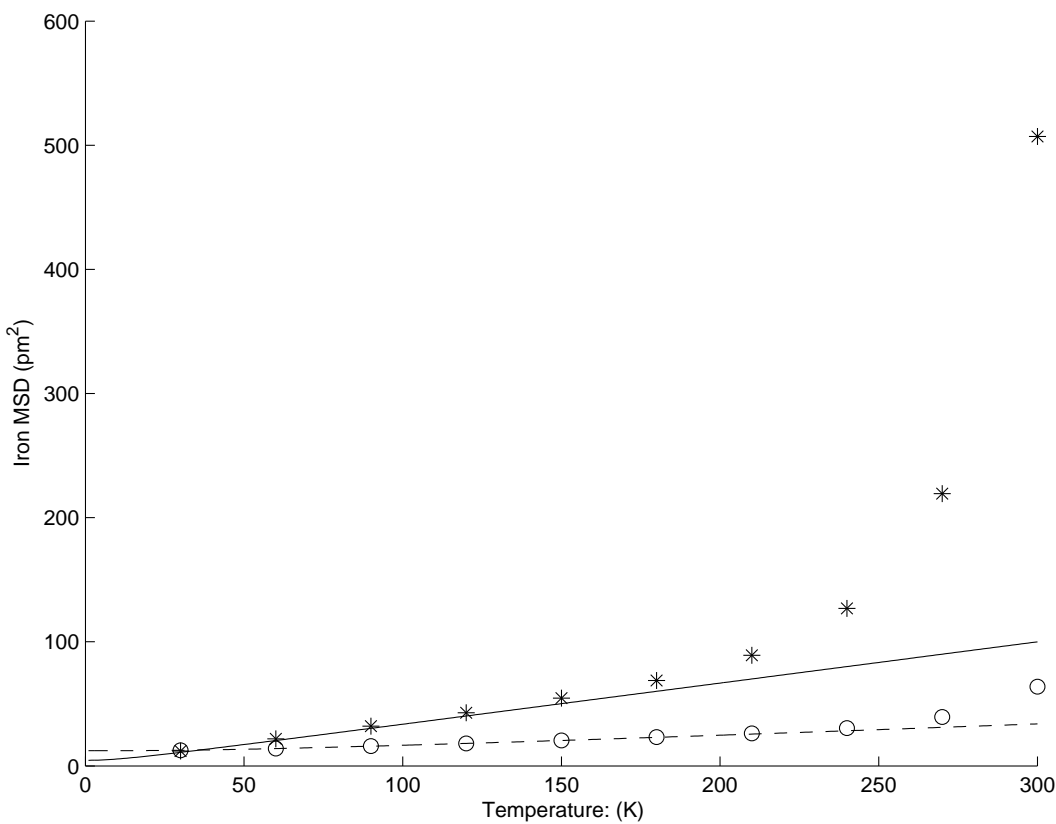}
\caption{ The cyt f iron in-plane and out-of-plane MSD from SCNMA.
Dashed line: iron in-plane MSD from classical NMA; Circle: iron
in-plane motion from SCNMA; Solid line: iron out-of-plane motion
from classical NMA; Star: iron out-of-plane motion from SCNMA}
\end{center}
\end{figure}\

\begin{table}[h]
\begin{center}
\includegraphics[width=14cm]{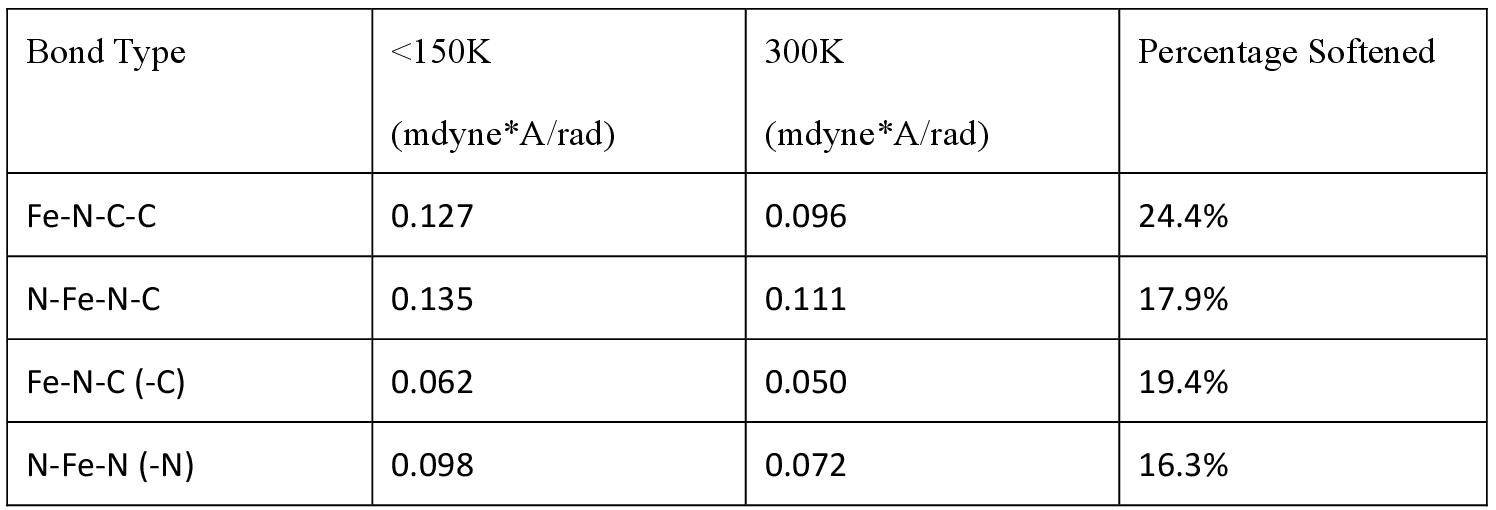}
\caption{ The softening of the iron dihedral force constant from
SCNMA}
\end{center}
\end{table}

\begin{figure}[h]
\begin{center}
\includegraphics[width=10cm]{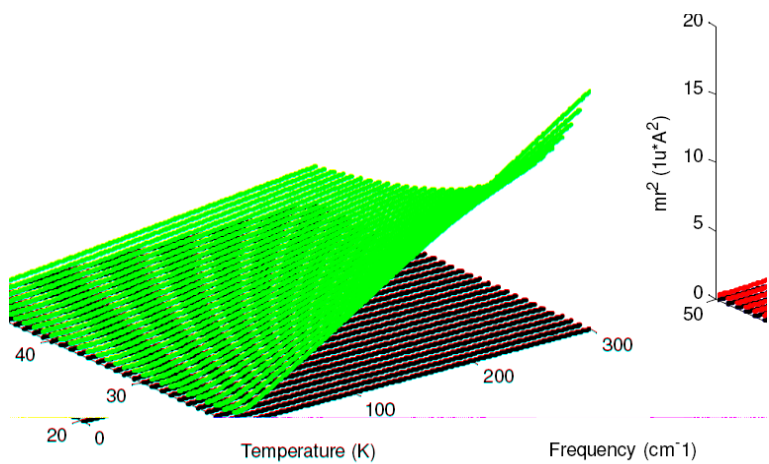}
\caption{ A plot of $<\sum_{i=1}^nm_ir_i^2>$ as a function of
temperature (T) and (low) frequency $\omega$ from equation (4)}
\end{center}
\end{figure}

\begin{figure}[h]
\begin{center}
\includegraphics[width=14cm]{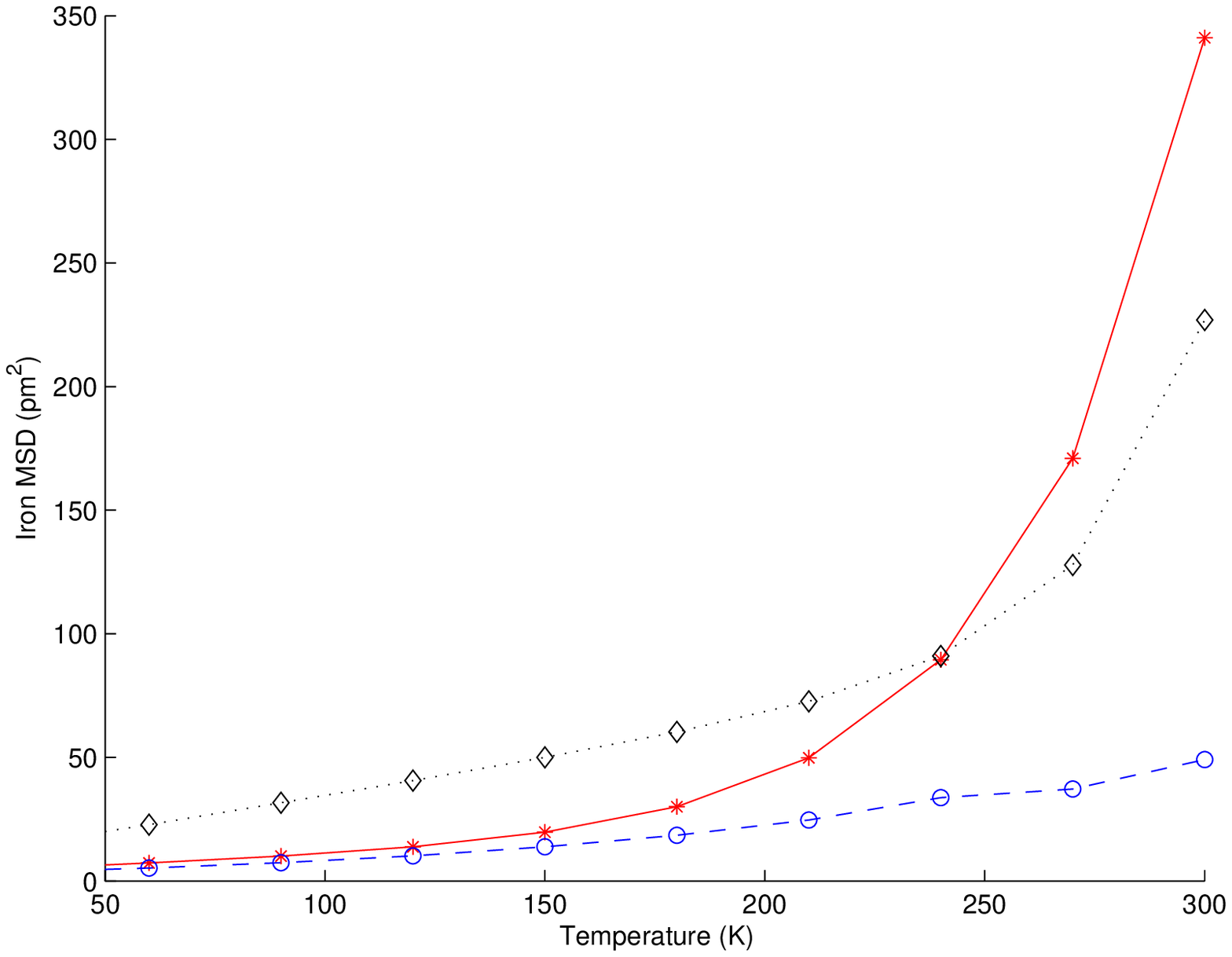}
\caption{Cyt f iron MSD from SCNMA in three frequency regimes: (a)
red star $<20cm^{-1}$(2.5meV),(b) black diamond:
$>20cm^{-1}$(2.5meV)and $<50cm^{-1}(6meV)$ and (c) blue circle
$>50cm^{-1}(6meV)$}
\end{center}
\end{figure}

Fig 3 and Fig 4 show the cyt f iron total MSD from SCNMA. These
results are in general agreement with the M\"ossbauer absorption
experimental results conducted on other heme proteins. The high
frequency ($>200cm^{-1}$) normal modes are softened on an average of
$1-2\%$. This is because the high frequency normal modes are
dominated by covalent stretching bonds which have relatively larger
strength and deeper potential wells. Moreover, these high frequency
atomic motions follow a strict, narrow Gaussian distribution. Fig 4
shows that the iron dynamical transition is caused by iron low
frequency out-of-plane motions. At lower frequency, the large iron
out-of-plane motion becomes possible because of the small energy
involved in changing the torsion angles. As temperature increases,
more and more displacement will spread out from the Gaussian
centroid. This low frequency classical behavior of the atomic
displacement distributions coincides with the fact that the
curvature of the potential function decreases over the distance from
the centroid, which results in the abrupt MSD increase seen in our
SCNMA model as compared with calculations implementing only Gaussian
distributions (Fig 4).

To analyze the protein flexibility, the force strength defined by
Zacca\"i \cite{Zaccai_force}is generally used by other authors
\cite{Bogdan_res}
\be k_0=\frac{k_B}{\frac{d<r^2>}{dT}}  \ee
From this definition,the iron force strength decreases by a factor
of $\sim$5-7. From NMA, the force constant is $k_0=m\omega^2$ and we
extract $\omega$ to find $r^2$. We found that the dihedral bonds,
which are the major dynamical element contributing to the iron
dynamical transition in our model, are softened by only $\sim20\%$,
as show in Table 1. The difference between the two definitions can
be explained with equation 1. A simple plot of equation 1 (Fig 5)
shows that $r^2$ increases exponentially below $50cm^{-1}$. Our
results show significant lowering of frequencies in this frequency
region.  Moreover, we found that atoms with internal coordinates
associated with soft bonds exhibit a larger MSD increase than
other atoms in one particular normal mode.

The MSD spread over frequency increases disproportionally upon
heating, as shown in Fig 6.  At temperatures below $\sim$150 K, the iron
MSD for the normal mode frequencies that are below $50cm^-1$ takes
about $84\%$ of the total iron MSD, while at 300K it increases to
$92\%$.

Generally speaking, the normal modes that participate in
biochemical reactions should have the largest motional amplitudes.
The largest amplitude among the iron out-of-plane normal
modes --- normally characterized as the "doming mode" --- has been
intensively studied experimentally \cite{doming_exp1}-\cite{doming_exp4} and
theoretically\cite{doming_theo1}-\cite{doming_theo4}.
This mode is Raman inactive In a four-fold symmetric porphyrin. The IR
spectroscopy and NRVS of cytochrome f failed to identify a well-resolved mode with
such character, and with the intensity expected for a heme doming
mode in the low frequency region. The modes around $~40cm^{-1}$ and
$~80cm^{-1}$ have been assigned to have the doming features by
various authors \cite{doming_exp4}\cite{doming_theo3}\cite{Bogdan_doming}.
In our SCNMA calculation, the normal modes $~80cm^{-1}$ have the features
of both iron doming motions and in-plane motions. The iron MSD in
the frequency regime $70-90cm^{-1}$ takes less than $10\%$ of the
total MSD, and these modes are close to Gaussian distributions at
room temperature from QHO theory. A theoretical study of the doming
mode has been carried out earlier by Li and Zgierski
\cite{doming_theo1} on a five-coordinated heme model. In the study,
the doming mode was predicted to be around $50cm^{-1}$ and was
calculated to be $35cm^{-1}$. Their analysis found that the doming
mode takes about $90\%$ of the iron MSD at room temperature. In one
previous NMA calculation, one $37cm^{-1}$ doming mode was found in
four-coordinate heme compound Fe(OEP), which takes $67\%$ of the
total iron MSD (unpublished results). In our six-coordinate cyt f
SCNMA, three normal modes that have the most iron MSD are
$(19cm^{-1}, 35cm^{-1}$ and $49cm^{-1})$ at low temperature and
softened to $(14cm^{-1}, 23cm^{-1}$ and $37cm^{-1})$. These three
modes take $63\%$ of the iron MSD and increase to $81\%$ at room
temperature. We assign them to the doming modes due to their
significant doming features. QHO theory indicates that these modes
are Gaussian distributions at low temperature $(<100K)$ and more
classical at room temperature (300K). As temperature increases,
these modes develop other features like saddling and ruffling due to
the softening of the dihedral bonds that are associated with these
modes. The energy distribution shows that these doming modes are
highly delocalized, i.e., the potential energy is distributed among
a large number of internal coordinates and the kinetic energy is
distributed among a large number of atoms. We also observe that the
iron low frequency motions are in phase with some other soft bond
atoms.

Besides the doming mode, some other significant water-protein
motions are observed in the frequency regime below $50cm^{-1}$.
These modes are softened by $20-50\%$ from low temperature to room
temperature. These results can also qualitatively explain the two
onsets of anharmonicity suggested by several authors
\cite{Doster_discussion} \cite{Cornicchi} as they proposed there are
two motional components: one happens at T~100K and one at
T~200-230K. As shown in Fig 6, lower frequency modes have a
relatively lower dynamical transition temperature.

The statistical properties of fast hydrated protein motions have
been analyzed by neutron scattering \cite{Doster_gaussian1} and
X-ray diffraction experiments\cite{Ringe}. At temperatures below
~200K, the displacement distribution is statistically a Gaussian.
However,a deviation from a Gaussian distribution becomes significant
at temperatures above ~240K. In our SCNMA calculation, below 100K,
the motions of individual atoms exhibit Gaussian behavior, but
starting from ~100 K, the atoms participating in soft internal
coordinates transition from Gaussian to classical distribution upon
heating. The percentage of heavy atoms that exhibiting classical
behavior rises to $~20\%$ at 300K. This result agrees with the
proposal by other authors who suggest the protein dynamical
transition is caused by water induced torsional
jump\cite{Doster_gaussian1}\cite{Zaccai1}. Furthermore, we also
quantitatively identify that the normal modes that contribute to the
dynamical transition lie in the frequency regime of $<50cm^{-1}$ at
temperatures below the dynamical transition temperature.

\section {Conclusion}
SCNMA can be used to study temperature dependent protein
vibrational motions. In the past, all such calculations assumed
Gaussian displacement distributions. However, single oscillators depart from
Gaussian distribution at higher temperatures.  This departure from
Gaussian behavior was studied quantitatively here using QHO theory and
SCNMA. Our study of heme c type cytochrome f has led us to identify
some specific features of the atomic interactions which may be of
general validity. Our results show that only a few normal modes account for most of the motional amplitudes of a significant set of bonds.
These modes lie in the frequency regime $<50cm^{-1} (or <6 meV
or >0.6ps)$. The higher frequency normal modes essentially
maintain a narrow Gaussian distribution. Above ~100K, the low
frequency modes transition from Gaussian to more classical
distributions upon heating, facilitating the softening of
dihedral (torsional) bonds, which seems to lead to the dynamical
transition.

\end{document}